\documentclass[12pt,a4paper]{article}

\usepackage{indentfirst}
\usepackage{graphicx}
\usepackage{amsmath}
\usepackage{bm}

\begin{document}


\title{\textbf{\large Atomic Parameters for the $2p^53p~^2[3/2]_2 - 2p^53s~^2[3/2]^o_2$ Transition of Ne I relevant in nuclear physics}}

\author{\normalsize Jiguang Li$^a$ \footnote{Li\_Jiguang@iapcm.ac.cn}, Michel Godefroid$^b$, Jianguo Wang$^a$ \medskip\\ 
{\small $^a$ Data Center for High Energy Density Physics,}\\ {\small Institute of Applied Physics and Computational Mathematics, Beijing 100088, China} \\
{\small $^b$ Chimie Quantique et Photophsique, Universit\'e libre de Bruxelles, B-1050 Brussels, Belgium}}

\date{\today}

\maketitle

\begin{abstract}
We calculated the magnetic dipole hyperfine interaction constants and the electric field gradients of $2p^53p~^2[3/2]_2$ and $2p^53s~^2[3/2]^o_2$ levels of Ne I by using the multiconfiguration Dirac-Hartree-Fock method. The electronic factors contributing to the isotope shifts were also estimated for the $\lambda = 614.5$ nm transition connecting these two states. Electron correlation and relativistic effects including the Breit interaction were investigated in details. Combining with recent measurements, we extracted the nuclear quadrupole moment values for $^{20}$Ne and $^{23}$Ne with a smaller uncertainty than the current available data. Isotope shifts in the $2p^53p~^2[3/2]_2 - 2p^53s~^2[3/2]^o_2$ transition based on the present calculated field- and mass-shift parameters are in good agreement with the experimental values. However, the field shifts in this transition are two or three orders of magnitude smaller than the mass shifts, making rather difficult to deduce changes in nuclear charge mean square radii. According to our theoretical predictions, we suggest to use instead transitions connecting levels arising from the $2p^53s$ configuration to the ground state, for which the normal mass shift and specific mass shift contributions counteract each other, producing relatively small mass shifts that are only one order of magnitude larger than relatively large field shifts, especially for the $2p^53s~^2[1/2]^o_1 - 2p^6~^1S_0$ transition. 

\vspace{0.5cm}
\end{abstract}

\section{Introduction}
Nuclear properties, especially for nuclei in the vicinity of the proton or neutron drip line, are of interest because the nuclear structure may be affected by the low binding energy of individual nucleons. Valuable information on the structure of nuclei can be obtained from nuclear electromagnetic moments and charge mean square radii. These physical quantities are accessible through the determination of hyperfine structures in atomic levels and isotope shifts in transition frequencies, if accurate atomic parameters are available~\cite{Kluge10}. 

It was suggested by early experimental studies that there exists two-proton halo structure for the proton drip-line nucleus $^{17}$Ne\cite{Ozawa94}. To confirm this speculation, many investigations have been stimulated to be performed~\cite{Geithner08, Tanaka10}. However, the evidence for $^{17}$Ne was not conclusive due to conflicting theoretical and experimental results. Experiments were carried out at ISOLDE/CERN to measure hyperfine structures and isotope shifts of the transition $2p^53s~^2[3/2]^o_2 -- 2p^53p~^2[3/2]_2$ ($\lambda=614.5$ nm) of $^{17}$Ne and other unstable neon isotopes up to the neutron-rich $^{28}$Ne by using fast-beam collinear laser spectroscopy~\cite{Geithner00, Geithner05, Geithner08, Marinova11}. The magnetic dipole and the electric quadrupole moments and the charge radii of the nuclei were deduced from these measurements in combination with atomic parameters. Nevertheless, the latter were obtained either by empirical methods~\cite{Ducas72, Geithner05, Klein96, Marinova11} or by \textit{ab initio} calculations but with some approximations~\cite{Sundholm92}.

In this work, we calculated the atomic parameters related to hyperfine interaction constants of $2p^53p~^2[3/2]_2$ and $2p^53s~^2[3/2]^o_2$ levels and isotope shifts of the transition between these two states using the multiconfiguration Dirac-Hartree-Fock (MCDHF) method. With assistance of the perturbation theory, a configuration space was constructed based on the active space approach. The present atomic parameters allowed us to extract the nuclear electric quadrupole moment values and to reproduce satisfactorily the experimental isotope shifts.

\section{Theory}
\subsection{MCDHF method}
The MCDHF method is written up in the monograph by Grant~\cite{Grant07}. Starting from the Dirac-Coulomb Hamiltonian
\begin{equation}
\label{Hamil}
H_{DC} = \sum_{i} \left[c {\bm \alpha}_i \cdot {\bm p}_i +(\beta_i -1)c^2 + V^N_i \right] + \sum_{i>j} 1/r_{ij},
\end{equation}
where ${\bm \alpha}$ and $\beta$ are Dirac matrices, and $V^N$ is the monopole part of the electron-nucleus Coulomb interaction. The atomic state functions (ASFs) approximating the eigenfunctions of the Hamiltonian (\ref{Hamil}) are obtained as linear combinations of symmetry-adapted configuration state functions (CSFs) with same parity $P$, total angular momentum $J$, and its component along $z$ direction $M_J$
\begin{equation}
\Psi(PJM_J) = \sum_{k=1}^{NCSF} c_k \Phi(\gamma_k PJM_J).
\end{equation} 
Here, NCSF is the number of CSFs, $\{c_k\}$ are the mixing coefficients and $\gamma_k$ represents in compact form the orbital occupation numbers, coupling trees and complementary labels required to uniquely define each CSF. Configuration state functions are formed by products of one-electron Dirac orbitals. In the self-consistent field (SCF) procedure, both the radial parts of the large and small component of the Dirac orbitals and the mixing coefficients are optimized to minimize the energies of the ASFs concerned. MCDHF calculations can be performed not only for a single level, but also for a portion of a spectrum in an extended optimal level (EOL) scheme where optimization is applied on a weighted sum of energies. The relativistic configuration interaction (RCI) method in which only the mixing coefficients are varied for a given fixed set of orbitals is often used to avoid convergence problems encountered in the SCF step or to investigate the effect of higher-order excitations. Additionally, the Breit interaction
\begin{equation}
B_{ij} = - \frac{1}{2r_{ij}} \left[ {\bm \alpha}_i \cdot {\bm \alpha}_j + \frac{({\bm \alpha}_i \cdot {\bm r}_{ij})({\bm \alpha}_j \cdot {\bm r}_{ij})}{r^2_{ij}} \right]
\end{equation} 
can also be included in this step.

\subsection{Hyperfine Interaction Constants}
The Hamiltonian of hyperfine interactions between the electrons and the electromagnetic multipole moments of the nucleus is expressed as a multipole expansion,
\begin{equation}
H_{hfs} = \sum_{k \geq 1} \mathbf{T}^{(k)} \cdot \mathbf{M}^{(k)},
\end{equation} 
where $\mathbf{T}^{(k)}$ and $\mathbf{M}^{(k)}$ are spherical tensor operators of rank $k$ in the electronic and the nuclear spaces, respectively~\cite{Schwartz55}. The $k=1$ term of the expansion represents the magnetic dipole interaction and the $k=2$ term the electric quadrupole interaction. According to perturbation theory, hyperfine structures corrected to the first order are described by diagonal magnetic dipole and electric quadrupole hyperfine interaction constants which are written as
\begin{align}
A_J & = \frac{\mu_I}{I} \frac{1}{[J(J+1)]^{1/2}} \langle \Psi(PJ) || \mathbf{T}^{(1)} || \Psi(PJ) \rangle, \\
\label{hfs-B} B_J & = 2Q_I \left[ \frac{J(2J-1)}{(J+1)(2J+3)} \right]^{1/2} \langle \Psi(PJ) || \mathbf{T}^{(2)} || \Psi(PJ) \rangle.  
\end{align}
$\mu_I$ and $Q_I$ are the magnetic dipole and the electric quadrupole moments of the nucleus with angular momentum $I$ and are defined through the relations 
\begin{align}
\mu_I &= \langle I M_I(=I) | M^{(1)}_0 | I M_I(=I) \rangle, \\
Q_I &= \langle I M_I(=I) | M^{(2)}_0 | I M_I(=I) \rangle.
\end{align}
The electronic tensor operators are sums of one-electron operators $\mathbf{t}^{(k)}$
\begin{align}
\mathbf{T}^{(1)} &= \sum_{j=1}^{N} \mathbf{t}^{(1)}(j) = \sum_{j=1}^{N} -i\alpha \left( {\bm \alpha}_j \cdot \mathbf{l}_j \mathbf{C}^{(1)}(j) \right) r_j^{-2}, \\
\mathbf{T}^{(2)} &= \sum_{j=1}^{N} \mathbf{t}^{(2)}(j) = \sum_{j=1}^{N} -\mathbf{C}^{(2)}(j) r_j^{-3}. 
\end{align}
Here, $i=\sqrt{-1}$ is the imaginary unit, $\alpha$ is the fine-structure constant, $\mathbf{l}$ is the orbital angular momentum operator and $\mathbf{C}^{(k)}$ is a spherical tensor. 

\subsection{Isotope Shifts}
The isotope shift (IS) for an atomic energy level is composed of the field shift (FS) and the mass shift (MS). The field shift, arising from the difference in the charge distribution between two isotopes with mass number $A$ and $A'$ $(A>A')$, is given in the approximation of the first-order perturbation theory by~\cite{Blundell87, Torbohm85}
\begin{equation}
\Delta E_{FS}^{A,A'} = \langle \Psi (PJM) | \sum_i \delta V^{N,AA'}_i | \Psi(PJM) \rangle.
\end{equation}
Here, $\delta V^{N,AA'} = V^{N,A} - V^{N,A'}$ and the nuclear potential $V^{N}$ for each isotope is produced by a two-parameter Fermi nuclear model~\cite{Parpia92, Li12}. Neglecting the higher-order nuclear moments~\cite{Seltzer69}, Eq.(11) is further simplified to
\begin{equation}
\Delta E_{FS}^{A,A'} = F \delta \langle r^2 \rangle^{A,A'},
\end{equation}
where
\begin{equation}
F = \frac{2\pi}{3} \left( \frac{Ze^2}{4\pi\epsilon_0} \right)  |\Psi(\mathbf{0})|^2
\end{equation}
is the field-shift factor proportional to the electronic total probability density at the origin, and $\delta \langle r^2 \rangle^{A,A'}$ is the difference of the nuclear charge mean square radius between these two isotopes. The mass shift between two isotopes $A$ and $A'$, caused by the motion of nucleus with the finite mass, is expressed as~\cite{Palmer87, Gaidamauskas11}
\begin{equation}
\Delta E_{MS}^{A,A'} = \frac{M'-M}{MM'} K_{MS}.
\end{equation}
Here, $M$ and $M'$ are the nuclear masses for isotopes $A$ and $A'$, respectively. The electronic factor $K_{MS}$ is defined by
\begin{equation}
\frac{K_{MS}}{M} \equiv \langle \Psi(PJM) | H_{MS} | \Psi(PJM) \rangle,
\end{equation}
where 
\begin{equation}
\label{MSO}
H_{MS} = \frac{1}{2M} \sum_{i,j} {\bm p}_i \cdot {\bm p}_j.
\end{equation}
The mass shift operator (Eq. (\ref{MSO})) can be split into two parts, that is, the one-body and the two-body mass shift operators
\begin{align}
H_{NMS} &= \frac{1}{2M} \sum_i p_i^2 \\
H_{SMS} &= \frac{1}{2M} \sum_{i \neq j} {\bm p}_i \cdot {\bm p}_j,
\end{align} 
which are also called the normal mass shift (NMS) and the specific mass shift (SMS) operator, respectively. It should be emphasized that the relativistic correction to the mass shift operator~\cite{Palmer87, Shabaev94, Tupitsyn03, Gaidamauskas11, Li12} are rather small (less than 1\%) for Ne I and were systematically omitted in the present work.

\section{Computational Model}
As mentioned earlier, hyperfine structures and isotope shifts are related not only to electronic factors in an atomic system, but also to nuclear parameters such as the nuclear electric quadrupole moment and charge mean square radii. Therefore, these nuclear parameters can be extracted from experimental values if accurate electronic factors are available from theory. In the framework of the MCDHF method, reasonable configuration spaces must be constructed for obtaining reliable electronic factors with enough precision. In this work, we adopted the active space approach to form the ASFs. A complete description of the computational model can be found in \cite{Li12} where we reported accurate transition energies and probabilities from levels in the first excited configuration to the ground state for Ne I.

We started from constructing the configuration space to account for the first-order electron correlation effects between valence orbitals, i.e. $2s$, $2p$, $3s$ and $3p$, in which CSFs were generated by replacing one or two orbitals occupied in the reference configurations with virtual orbitals. The reference configurations are $\{2s^22p^53s\}$ for the lower (odd) level and $\{2s^22p^53p\}$ for the upper (even) one, respectively. Virtual orbitals were augmented layer by layer in order to monitor the convergence trend of the physical quantities under investigation and to further estimate the uncertainties in their evaluation. 
For the first four correlation layers labelled by $n$ ($n=3, 4, 5, 6$) all possible orbitals ($l \le n-1$) with angular symmetries up to $l = 5$ were included, while the angular momentum of the virtual orbitals was limited to $l=3$ for the three extra layers ($7 \le n \le 9$). As a result, we marked the computational model for the former and the latter as $n$SD and $nf$SD, respectively.  Self-consistent field (SCF) calculations were performed in which the occupied orbitals of each reference configuration were treated as spectroscopic and kept frozen in the subsequent steps. To avoid convergence problems in SCF calculations, virtual orbitals only in the added layer were optimized. The first-order correlation involving the $1s$ core electrons was further taken into account through the RCI computations where CSFs generated by SD substitutions from all occupied orbitals in the reference configurations to the largest virtual orbitals set were included. These calculations were labelled ``CC" in the tables.

Higher-order correlation corrections, which cannot be accounted for by SD-excitation CSFs from the monoreference configuration, are more difficult to deal with, since the number of CSFs grows rapidly and easily goes beyond the current computer resources. Alternatively, we adopted a single- and double-multireference (MR) approach by selecting the dominant CSFs from the first-order configuration space and adding them to the multireference sets. The latter are 
\begin{equation}
\{2s^22p^53s, 2s^22p^33s3p^2, 2s2p^53s3d, 2s^22p^33s3d^2\}
\end{equation}
and 
\begin{equation}
\{2s^22p^53p, 2s^22p^33p3d^2, 2s^22p^33p4p^2, 2s2p^53p3d, 2s2p^43s3p4p\}
\end{equation}
for the lower odd and upper even states. The $1s$-core SD excitations were restricted to those already included in the monoreference CC calculations and the added SD excitations were limited to the ones from the valence orbitals of the multi-reference configuration sets to the virtual orbital set $\{4s, 4p, 4d, 4f\}$. It means that we only considered the higher-order correlation effects between valence orbitals. The configuration spaces were marked with ``MR-4SD". At last, the Breit interaction was included through the RCI calculations in the latest configuration spaces. In practical, these calculations were performed by using the GRASP2K package~\cite{GRASP}.

\section{Results and Discussion}
\subsection{Nuclear Electric Quadrupole Moments of $^{21,23}$Ne}
Table~\ref{hfs} reports the magnetic dipole hyperfine interaction constants and the electric field gradient (EFG) $q = \langle \Psi(PJ) || {\bf T}^{(2)} || \Psi(PJ) \rangle$ at the nucleus, as functions of the computational model, for $2p^53s~^2[3/2]^o_{2}$, and $2p^53p~^2[3/2]_2$ levels of Ne I. As can be seen from this table, the magnetic dipole hyperfine interaction constant $A$ is more sensitive to electron correlation than the electric field gradient $q$. Moreover, excitations involving the $1s$ core electron significantly contribute to both physical quantities. In addition, the effect of the Breit interaction is remarkable in the $2p^53p~^2[3/2]_2$ level, while tiny in the $2p^53s~^2[3/2]^o_2$. For the magnetic dipole hyperfine interaction constants, the present results differ from previous experimental values~\cite{Geithner05, Grosof58} by less than 2\% and 3\% for the lower and upper levels, respectively. Furthermore, our values of the electric field gradient for $2p^53s~^2[3/2]^o_2$ is in excellent agreement with the results of Sundholm and Olsen~\cite{Sundholm92} using the finite-element multiconfiguration Hartree-Fock method. We should stress, however, that the electron correlation involving $1s$ shell was not considered in their work. Moreover, the relativistic corrections to the EFG was estimated by performing limited qusi-relativistic CI in the valence shells.    

Using our electric field gradient value ($q=-4.681$), we extracted the nuclear quadrupole moments for $^{21}$Ne and $^{23}$Ne, respectively, from the most recent measurements of the electric quadurpole hyperfine interaction constants~\cite{Geithner05}, using the following relation
\begin{equation}
{\rm Q} = \frac{B}{234.9647 q}
\end{equation}
in which $\rm Q$, $B$ and $q$ are expressed in barns, MHz and a.u., respectively. The results are displayed in Table~\ref{NQM} as well as other values available~\cite{Grosof58, Ducas72, Sundholm92, Geithner05, DeRydt13}. Among these, Q values for $^{21}$Ne reported by Grosof \textit{et al.}~\cite{Grosof58} and Ducas \textit{et al.}~\cite{Ducas72} were obtained by fitting the measured hyperfine splittings. It can be seen that the corresponding values have relatively large errors due to the approximations made in the fitting procedures. Note that the value Q($^{21}$Ne) = 103(8) was selected by Stone~\cite{Stone05} in the compilation of nuclear magnetic dipole and electric quadrupole moments. Using the experimental $B$-factor measured by Grosof \textit{et al.}, Sundholm and Olsen~\cite{Sundholm92} extracted Q($^{21}$Ne) using their electric field gradient value, which was retained in the latest compilation of Pyykk\"o~\cite{Pyykko08}. Although the agreement with our value is very good, the uncertainty of the present Q value for $^{21}$Ne (about 0.5 \%) is likely to be smaller than the one by Sundholm and Olsen~\cite{Sundholm92}, thanks to a better description of electron correlation and relativity. In a more recent experiment, Geithner \textit{et al.} measured the electric quadrupole hyperfine interaction constant of $^{23}$Ne for the first time in a collinear laser spectroscopy experiment~\cite{Geithner05}. Applying the relation ${\rm Q}^A/{\rm Q}^{A'} = B^A/B^{A'}$ between two isotopes $A$ and $A'$ and using the Q($^{21}$Ne)=102.9(7.5) value obtained by Ducas \textit{et al.}~\cite{Ducas72}, they deduced the nuclear electric quadrupole moment of $^{23}$Ne with a $\simeq 4\%$ uncertainty. Furthermore, De Rydt~\cite{DeRydt13} reevaluated the value in the same way but using a weight mean factor of ${\rm Q}^{23}/{\rm Q}^{21} = 1.407(41)$, which reduced the error bars to about 3\%. These two values are in good agreement with our \textit{ab initio} calculation. We should point out that the uncertainty in our Q($^{23}$Ne) value mainly arises from the experimental error in the $B$ factor reported by Geithner \textit{et al.}~\cite{Geithner05}.

\begin{table}[!ht]
\caption{\label{hfs} Magnetic dipole hyperfine interaction constants $A$ (in MHz) and electric field gradients $q$ (in a.u.) for $2p^53s~^2[3/2]^o_{2}$ and $2p^53p~^2[3/2]_2$ levels of $^{21}$Ne. $I=3/2$ and $\mu_I$ = -0.66171797 n.m.~\cite{Stone05} were used to calculate $A$ values.}
\begin{tabular}{ccccccc}
\hline
\hline
 & \multicolumn{2}{c}{$2p^53s~^2[3/2]^o_2$} && \multicolumn{2}{c}{$2p^53p~^2[3/2]_2$}  \\
\cline{2-3}\cline{5-6} Model & $A$ & $q$ && $A$ & $q$  \\
\hline
DF          &   -248.72   & -4.978 && -179.65   & -2.711   \\
3SD        &   -248.18   & -4.726 && -192.77   & -2.708  \\
4SD        &   -302.62   & -4.783 && -184.30   & -2.448  \\
5SD        &   -289.46   & -4.748 && -187.99   & -2.543  \\
6SD        &   -296.19   & -4.749 && -187.20   & -2.546  \\
7fSD       &   -293.93   & -4.749 && -187.41   & -2.564  \\
8fSD       &   -295.40   & -4.745 && -187.69   & -2.583  \\
9fSD       &   -295.31   & -4.745 && -187.57   & -2.586  \\
CC          &   -258.50   & -4.704 && -176.71   & -2.479  \\
MR-4SD  &   -262.34   & -4.684 && -175.92   & -2.475 \\
$+$ Breit        &   -262.89   & -4.681 && -180.72   & -2.676  \\[0.2cm]
Grosof \textit{et al.}~\cite{Grosof58}        & -267.68(3)      &                   &&                     &     \\
Geithner \textit{et al.}~\cite{Geithner05}  & -267.62(15)    &                   && -185.83(21) &      \\
Sundholm and Olsen~\cite{Sundholm92} &                       &  -4.675(30) &                       & &   \\
\hline
\hline
\end{tabular}
\end{table}
 
\begin{table}[!ht]
\centering
\caption{\label{NQM} Nuclear electric quadrupole moments Q (in mb) of $^{21,23}$Ne isotopes.}
\begin{tabular}{cccc}
\hline
\hline
                                                                  &  Q($^{21}$Ne)   & Q($^{23}$Ne)  \\
\hline
This work                                                   &  101.44(20)       & 141.8(6.0)         \\
Grosof \textit{et al.}~\cite{Grosof58}          &    93(10)            &                         \\
Ducas \textit{et al.}~\cite{Ducas72}           &  102.9(7.5)        &                         \\
Sundholm and Olsen~\cite{Sundholm92} &  101.55(75)       &                          \\
Geithner \textit{et al.}~\cite{Geithner05}   &                           & 145(6)              \\
De Rydt \textit{et al}~\cite{DeRydt13}       &                           & 142.9(4.3)          \\
\hline
\hline
\end{tabular}
\end{table}

\subsection{Isotope Shift in the $2p^53s~^2[3/2]^o_2 -- 2p^53p~^2[3/2]_2$ transition}
Isotope shift parameters, including normal mass shift $\Delta \tilde{K}^{NMS}$, specific mass shift $\Delta \tilde{K}^{SMS}$ and field shift $F$ factors, are given in Table~\ref{IS-Fact} for the $2p^53s~^2[3/2]^o_2 -- 2p^53p~^2[3/2]_2$ transition in Ne I. It was found that electron correlation effects on the specific mass shift factor $\Delta \tilde{K}^{SMS}$ are huge in the case under investigation. One indeed observes with this respect that the sign of the Dirac-Fock $\Delta \tilde{K}^{SMS}$ value changes when including electron correlation. When taking the $1s$-core excitations into account,  the SMS parameter increased by a factor of 6. This result is, however, surprising in view of the inconsistency with other theoretical~\cite{Ahmad85} or experimental~\cite{Konz92, Basar97} results. This is an artefact and is most likely due to the unbalance description of core correlation between the two levels concerned. This effect is actually rather small on the level shift, but large enough to manifest itself in such a tiny transition specific mass shifts. To offset this effect, higher-order correlations involving the 1s core should be accounted for, but these calculations are definitely beyond the limits in current computational resources. For eliminating the $1s$-core excitations unbalanced contributions, we subtracted the 251 GHz u core-correlation (CC $-$ 9fSD) differences in the specific mass shift factors, from the ``MR-4SD" and ``Breit" values. The so-corrected $\Delta \tilde{K}^{SMS}$ parameters, marked by an asterisk in Table~\ref{IS-Fact}, are in reasonably good agreement with the other available results.    

Compared with the specific mass shift factor, the normal mass shift and the field shift parameters were stable with expanding the configuration space. Instead of an \textit{ab initio} calculation, the normal mass shift factor can be estimated in the nonrelativisitc approximation from the scaling law $\Delta \tilde{K}^{NMS} = - \nu/1823$ where $\nu$ is transition frequency~\cite{Godefroid01}. Using the experimental transition frequency~\cite{NIST}, one obtains a value of $267.04$ GHz u that agrees with our \textit{ab initio} calculation. For the field-shift factor,  Marinova \textit{et al.} obtained $F=40(4)$ MHz/fm$^2$ applying the semiempirical approach based on the Goudsmit-Fermi-Segr\'e formula. This value is consistent with our result. 
 
Using the present calculated mass- and field-shift parameters, we obtained the isotope shifts between $^{A}$Ne and $^{20}$Ne that are reported in Table~\ref{IS}. The changes in mass and nuclear mean square charge radii were taken from Ref.~\cite{Geithner08}. As can be seen from this table, our isotope shifts are in good agreement with experimental values, and the difference is less than 4\%.    

Field shifts in light elements such as Ne ($Z=10$) are so small that it is difficult to extract nuclear charge radii from isotope shifts. Especially for the $2p^53p~^2[3/2]_2 - 2p^53s~^2[3/2]^o_2$ transition, the field-shift factor is only 32.3 MHz/fm$^2$, producing a field shift of only 0.549 MHz between $^{19}$Ne and $^{20}$Ne. Moreover, the mass shifts for this transition are two or three orders of magnitude larger than the field shifts. In our opinion, this transition is therefore not suitable for deducing reliable nuclear charge radii. Alternatively, we found from the present calculations (see Table~\ref{IS-II}) that the transition from levels of the $2p^53s$ configuration to the $2p^6~^1S_0$ ground state are much better candidates for this purpose. The reason is that field shifts are larger for $3s - 2p$ than for $3s-3p$ transitions while mass shifts are relatively small, especially for the $2p^53s~^2[1/2] - 2p^6~^1S_0$ transition for which the specific mass shift contribution counteracts the normal mass shift. Therefore, it would be worthwhile to measure the isotope shifts in these transitions in order to obtain more accurate nuclear charge radii for Ne isotopes.
 
\begin{table}[!ht]
\centering
\caption{\label{IS-Fact} Mass-shift $\Delta \tilde{K}^{MS}$ (in GHz u) and Field-shift $F$ (in MHz/fm$^2$) parameters for the $2p^53s~^2[3/2]^o_2 - 2p^53p~^2[3/2]_2$ transition in Ne I. Normal mass shift (NMS) $\Delta \tilde{K}^{NMS}$ (in GHz u) and specific mass shift (SMS) $\Delta \tilde{K}^{SMS}$ (in GHz u) factors were also displayed.}
\begin{tabular}{cccccc}
\hline
\hline
       &          \multicolumn{3}{c}{$\Delta \tilde{K}$}                                 &&       \\
\cline{2-4} Model  &   $\Delta \tilde{K}^{NMS}$ & $\Delta \tilde{K}^{SMS}$ & $\Delta \tilde{K}^{MS}$ &&  $F$  \\
\hline
DF          &     1643     &    -648     &       995        &&  35.7 \\
3SD        &     331       &    -761     &      -430        &&  30.7 \\
4SD        &     350       &    435      &       785        &&  26.1 \\
5SD        &     153       &    86        &       239        &&  31.7 \\
6SD        &     204       &    47        &       251        &&  29.7 \\
7fSD       &     201       &    63        &       263        &&  29.9 \\
8fSD       &     218       &    49        &       267        &&  30.3 \\
9fSD       &     215       &    46        &       261        &&  30.1 \\
CC          &     231       &    297      &       528        &&  30.2 \\
MR-4SD &     266       &    349      &       615        &&  31.6 \\
               &                  &      98*     &       364*      &&          \\
$+$ Breit        &     275       &    352      &       627       &&  32.3 \\
               &                  &    101*     &       376*      &&          \\       [0.2cm]
\multicolumn{6}{c}{Others}  \\
Ref.~\cite{Ahmad85}    &   &  92         &              &&            \\
Ref.~\cite{Konz92}       &   & 104(0.6) &              &&            \\   
Ref.~\cite{Basar97}      &   & 97(1)      &              &&            \\
Ref.~\cite{Marinova11} &    &              &              &&   40(4) \\
\hline
\hline
\end{tabular}
\end{table}

\begin{table}[!ht]
\centering
\caption{\label{IS} $2p^53s~^2[3/2]^o_2 - 2p^53p~^2[3/2]_2$ transition isotope shift (IS) (in MHz) relative to $^{20}$Ne. Mass differences ($\Delta^{A,20} = \frac{1}{M(^{20}Ne)} - \frac{1}{M(^{A}Ne)}$) (in 1/u) and nuclear mean-square charge radii ($\langle r^2 \rangle^{A,20} = \langle r^2 \rangle^{A} - \langle r^2 \rangle^{20}$) (in fm$^2$) between given isotope A and $^{20}$Ne were calculated with data presented in Ref.\cite{Geithner08}.}
\begin{tabular}{ccccccc}
\hline
\hline
A  & $\Delta^{A,20}$ & $\delta \langle r^2 \rangle^{A,20}$ & MS & FS &  IS & Exp.~\cite{Geithner08} \\
\hline
17 & -0.008743390    &  0.220  & -3287              &   7.084     &   -3280        & -3183.3       \\
18 & -0.005519035    & -0.207  & -2075              &  -6.686     &   -2082        & -1995.5       \\
19 & -0.002607463    &  0.017  & -980                &   0.549     &   -980          & -947.39       \\
21 &  0.002385902    & -0.217  &  897                &  -7.009     &    890          &  874.94       \\
22 &  0.004546555    & -0.321  &  1710              &  -10.366   &    1699        &  1663.58      \\
\hline
\hline
\end{tabular}
\end{table}

\begin{table}
\centering
\caption{\label{IS-II} Mass-shift $\Delta \tilde{K}^{MS}$ (in GHz u) and field-shift $F$ (in MHz/fm$^2$) parameters for transitions from states in the $2p^53s$ configuration to the ground state $2p^6~^1S_0$ in Ne I. Normal mass shift (NMS) $\Delta \tilde{K}^{NMS}$ (in GHz u) and specific mass shift (SMS) $\Delta \tilde{K}^{SMS}$ (in GHz u) factors were also displayed.}
\begin{tabular}{cccccc}
\hline
\hline
       &          \multicolumn{3}{c}{$\Delta \tilde{K}$}                                 &&       \\
\cline{2-4} Transition  &   $\Delta \tilde{K}^{NMS}$ & $\Delta \tilde{K}^{SMS}$ & $\Delta \tilde{K}^{MS}$ &&  $F$  \\
\hline
$2p^53s~^2[3/2]^o_2 - 2p^6~^1S_0$    &    -2255       &   2065       &    -190       && 138.11  \\
$2p^53s~^2[3/2]^o_1 - 2p^6~^1S_0$    &    -2270       &   2091       &    -179       && 137.13  \\
$2p^53s~^2[1/2]^o_0 - 2p^6~^1S_0$    &    -2295       &   2100       &    -195       && 138.34  \\
$2p^53s~^2[1/2]^o_1 - 2p^6~^1S_0$    &    -2269       &   2235       &    -34         && 137.13  \\
\hline
\hline
\end{tabular}
\end{table}

\section{Conclusion}
In this work, we reported the magnetic dipole hyperfine interaction constants and the electric field gradient at the nucleus of $2p^53p~^2[3/2]_2$ and $2p^53s~^2[3/2]^o_2$ levels and the electronic factors of isotope shifts in the transition between these two states for Ne~I. These results were obtained by combining the multiconfiguration Dirac-Hartree-Fock method with relativistic CI calculations and the active space approach to take electron correlation and relativistic effects into account. Using the calculated electric field gradient, we extracted the nuclear quadrupole moments for $^{21}$Ne and $^{23}$Ne. The present results have a better accuracy than the values available in the literature. Based on calculated mass- and field-shift factors, experimental isotope shifts for the transition concerned were satisfactorily reproduced. However, we found that this transition is not the best candidate for deducing changes in the nuclear charge mean square radii due to small field shifts. We therefore suggested for this purpose the use of transitions connecting the $2p^53s$ levels to the ground state $2p^6~^1S_0$, offering much larger field shifts, and smaller mass shifts due to strong cancellation between normal and specific mass shift contributions.   

\section*{Acknowledgement}
This work has been supported by National Natural Science Foundation of China under Grant Nos. 11404025 and 91536106, the China Postdoctoral Science Foundation under Grant No. 2014M560061, the Belgian F.R.S.-FNRS Fonds de la Recherche Scientifique (CDR J.0047.16) and the BriX IAP Research Program No. P7/12 (Belgium).


\end{document}